\documentclass[a4paper,12pt]{article}
\usepackage{amsmath,amsfonts,amssymb,amsthm,amstext,url}
\usepackage[cp1251]{inputenc}
\usepackage[T2A]{fontenc}
\usepackage[russian,english]{babel}
\begin{document}

\title{
A simple algorithm for\\
the evaluation of the hypergeometric series\\
using quasi-linear time and linear space}
\author{S.V.Yakhontov \\ \\
Saint Petersburg State University \\
Faculty of Mathematics and Mechanics \\
SergeyV.Yakhontov@gmail.com\\
June 14, 2011}
\date{}

\maketitle

\newtheorem*{notation}{Notation}
\newtheorem*{definition}{Definition}
\newtheorem*{proposition}{Proposition}
\newtheorem*{lemma}{Lemma}
\newtheorem*{theorem}{Theorem}
\newtheorem*{corollary}{Corollary}

\abstract {
A simple algorithm with time complexity
$O(M(n)\log(n)^2)$ and space complexity $O(n)$ for the evaluation
of the hypergeometric series with rational coefficients is constructed
($M(n)$ being the complexity of integer multiplication). It is shown that this
algorithm is suitable in practical informatics for constructive analogues
of often used constants of analysis.
}


\paragraph{Introduction.}
In this paper we construct an algorithm for the calculation of the approximate 
values of the hypergeometric series with rational coefficients
whose implementation
is simple and which is quasi-linear in time and linear in space on the machine \textit{Schonhage}
\cite{yakhontov:bib:Schonhage}.
Such series are used for the calculation of some mathematical constants
of analysis and of the values of elementary functions at rational points.

$Sch(FQLIN-TIME//LIN-SPACE)$ will be used to denote the class of algorithms
which are computable on Schonhage and are quasi-linear in time
and linear in space. The main feature of Schonhage is its ability
to execute recursive calls of procedures. Quasi-linear means that
the complexity function is bounded by $O(n\log(n)^k)$ for some $k$.

The main advantage of algorithms based on series expansions
is the relative simplicity of both the algorithms and the analysis of
their computational complexity. Besides we can compute all the most
commonly used constants of analysis using series expansions.
For calculations with a small number of digits after the binary (or decimal) point,
series are more efficient than other methods because of the small constants in
estimations of their computational complexity. Therefore such
algorithms are important in computer science for practical applications.

It is known \cite{yakhontov:bib:Haible} that linearly convergent
hypergeometric series with rational coefficients can be
calculated using the binary splitting method with time complexity
$O(M(n)(\log(n))^ 2)$ and space complexity $O(n\log(n))$ (where $M(n)$
denotes the complexity of multiplication of $n$-bit integers). In recent publications,
for example \cite{yakhontov:bib:Cheng}, algorithms based on a modified
binary splitting method for the evaluation of linearly convergent
hypergeometric series with time complexity $O(M(n)(\log(n))^2)$
and space complexity $O(n)$ are described.

In this paper we propose an algorithm for the evaluation of the hypergeometric
series which is simpler in its practical implementation than
the algorithm from \cite{yakhontov:bib:Cheng}; the proposed algorithm is
also quasi-linear in time and linear in space.
The idea of working with a given accuracy to calculate the value of
the hypergeometric series with a linear space complexity can be found at
http://numbers.computation.free.fr/Constants/Algorithms/ \\ splitting.html.

The computational complexity of the constructive real numbers and functions
is discussed in detail in \cite{yakhontov:bib:Ko}. The set of
constructive real numbers with quasi-linear time and linear
space complexity of calculating their dyadic approximations will
be denoted by $Sch(FQLIN-TIME//LIN-SPACE)_{CF}$ ($CF$ is the Cauchy function).

From now on, $n$ will denote the length of the record of accuracy
$2^{-n}$ of dyadic rational approximations. We will use $\log(k)$ for
logarithms base 2.


\paragraph{1. Binary splitting method.}
This method is used to calculate the values of linearly convergent
series with rational coefficients, in particular to calculate
the hypergeometric series of the form
\begin{align}
S=\sum_{i=0}^{\infty}{\frac{a(i)}{b(i)}
\prod_{j=0}^{i}{\frac{p(j)}{q(j)}}},
\label{eq:yah:SeriesDef}
\end{align}
where $a$, $b$, $p$, and $q$ are polynomials with integer coefficients;
this series is linearly convergent if its partial sum
\begin{align}
S(\mu(k))=\sum_{i=0}^{\mu(k)}{\frac{a(i)}{b(i)}
\prod_{j=0}^{i}{\frac{p(j)}{q(j)}}},
\label{eq:yah:SeriesPartialSum}
\end{align}
where $\mu(k)$ is a linear function of $k$, differs from the exact value
by not more than $2^{-k}$: $|S-S(\mu(k))|\le 2^{-k}$.

In its classical variant the binary splitting method works as follows.
Put $k_1=\mu(k)$. We consider the partial sum \eqref{eq:yah:SeriesPartialSum}
for some integers $i_1$ and $i_2$, $0\le i_1\le k_1$, $0\le i_2\le k_1$,
$i_1\le i_2 $:
\begin{align}
S(i_1,i_2)=\sum_{i=i_1}^{i_2}{\frac{a(i)p(i_1)
\ldots p(i)}{b(i)q(i_1)\ldots q(i)}}.
\label{eq:yah:Seriesi1i2Sum}
\end{align}
We calculate
\begin{align*}
&P(i_1,i_2)=p(i_1)\ldots p(i_2),\quad Q(i_1,i_2)=q(i_1)\ldots q(i_2),\\
&B(i_1,i_2)=b(i_1)\ldots b(i_2),\quad and\quad
T(i_1,i_2)=B(i_1,i_2)Q(i_1,i_2)S(i_1,i_2).
\end{align*}
If $i_1=i_2$, then
these values are calculated directly. Otherwise, the series is divided into
two parts, left and right, and $P(i_1,i_2) $,
$Q(i_1,i_2)$, and $B(i_1,i_2)$ are calculated for each
part recursively. Then the values obtained are combined:
\begin{align}
\begin{split}
P(i_1,i_2)&=P_l P_r, \quad Q(i_1,i_2)=Q_l Q_r, \quad B(i_1,i_2)=B_l B_r,\\
T(i_1,i_2)&=B_r Q_r T_l + B_l P_l T_r.
\end{split}
\label{eq:yah:PQBTCalculation}
\end{align}
The algorithm starts with $i_1=0$, $i_2=k_1$. After calculating
$T(0,k_1)$, $B(0,k_1)$, and $Q(0,k_1)$, we divide $T(0, k_1)$
by $B(0,k_1)Q(0,k_1)$ to get the result with the given accuracy.
The lengths of $T(0,k_1)$ and $B(0,k_1)Q(0,k_1)$ are proportional
to $k\log(k)$; therefore the binary splitting algorithm is quasi-linear
in space; the time complexity of this algorithm is $O(M(k)\log(k)^2)$
\cite{yakhontov:bib:Haible}.

\paragraph{2. The basic algorithm of class $Sch(FQLIN-TIME//LIN-SPACE)$.}
We will modify the binary splitting method for evaluation of the hypergeometric series
so that the algorithm is simple and is in class $Sch(FQLIN-TIME//LIN-SPACE)$.

Let's suppose that we want to calculate the values of the hypergeometric series
\eqref{eq:yah:SeriesDef} with an accuracy of $2^{-n}$. It's enough to
calculate the partial sum \eqref{eq:yah:SeriesPartialSum} with accuracy
$2^{-(n+1)}$ because
\begin{align*}
|S-S(\mu(n+1))^*|
&\le|S-S(\mu(n+1))|+|S(\mu(n+1))-S(\mu(n+1))^*|\\
&\le 2^{-(n+1)}+2^{-(n+1)}=2^{-n};
\end{align*}
here $S(\mu(n+1))^*$ is an approximate value of $S(\mu(n+1))$.
Put $r=\mu(n+1)$. Take the minimum value $k_1$ such that
$2^{k_1}\ge r$; let $r_1=\lceil\frac{r}{k_1}\rceil$. We write
the partial sum \eqref{eq:yah:SeriesPartialSum} as follows:
\begin{align}
P(r)
=\sigma_1+\tau_2[\sigma_2+\tau_3[\sigma_3+\ldots+
\tau_{k_1-1}[\sigma_{k_1-1}+\tau_{k_1}\sigma_{k_1}]]],
\label{eq:yah:PmukEquation}
\end{align}
where
\begin{align*}
\sigma_1&=
\sum_{i=0}^{r_1-1}{\frac{a(i)}{b(i)}
\prod_{j=0}^{i}{\frac{p(j)}{q(j)}}}=\sigma_1',\\
\tau_2&=
\prod_{j=0}^{r_1}{\frac{p(j)}{q(j)}},\quad
\sigma_2=
\frac{a(r_1)}{b(r_1)}+
\sum_{i=r_1+1}^{2r_1-1}{\frac{a(i)}{b(i)}
\prod_{j=r_1+1}^{i}{\frac{p(j)}{q(j)}}}
=\xi_2+\sigma_2',
\\
\tau_3&=
\prod_{j=r_1+1}^{2r_1}{\frac{p(j)}{q(j)}},\quad
\sigma_3=
\frac{a(2r_1)}{b(2r_1)}+
\sum_{i=2r_1+1}^{3r_1-1}{\frac{a(i)}{b(i)}
\prod_{j=2r_1+1}^{i}{\frac{p(j)}{q(j)}}}
=\xi_3+\sigma_3',
\\
\\
&\cdots
\end{align*}
The $\sigma_t'$ are calculated by the usual binary
splitting method for the sum \eqref{eq:yah:Seriesi1i2Sum},
where $i_1=(t-1)r_1+1$, $i_2=t\cdot r_1-1$; for $\sigma_1'$
we take $i_1=0$, $i_2=r_1-1$.

We introduce the notation
\begin{align*}
&\omega=l(W)+1,\thickspace W=\max(A,B),\\
&\text{where}\thickspace
A=\max_{i=0..r}(|a(i)|,|b(i)|),\thickspace
B=\max_{j=0..r}(|p(j)|,|q(j)|)
\end{align*}
(here, $l(u)$ is the length of the bit representation of $u$).
Let's obtain an estimate of the computational complexity of the calculation
of $\sigma_t$ and $\tau_t$.

\begin{lemma}[1]
The time complexity of the binary splitting method for calculation of
$\sigma_t$ is bounded by $O(M(r)\log(r))$; the space complexity is
bounded by $O(r)$.
\begin{proof}
Consider an arbitrary maximal chain of recursive calls
derived from calculations by Formulas \eqref{eq:yah:PQBTCalculation}.
Suppose that the numbering in the chain begins with its deepest
element: $i=1,\ldots,\varsigma$, where $\varsigma$ is the length of the chain,
$\varsigma\le\lceil\log(2r_1)\rceil$.

Using mathematical induction on $i$, we show that the length of
the representation of $T$ at level $i$ satisfies
$l(T_i)<2^{i+1}\omega+2^{i}$. Note that
$l(P_i)\le 2^{i}\omega$, $l(Q_i)\le 2^{i}\omega$,
and $l(B_i)\le 2^{i}\omega$ because there is a doubling of the length of the number
representation when $i$ increases. The induction begins with $i=1$:
$l(T_1)\le 2\omega<2^{2}\omega+2^{1}$; next, the induction step is
\begin{align*}
l(T_{i+1})
<2^{i}\omega+2^{i}\omega+2^{i+1}\omega+2^{i}+1
<2^{(i+1)+1}\omega+2^{i+1}.
\end{align*}
Note that, based on this inequality, $T_{\varsigma}$ has length
$O(r)$ because
\begin{align*}
l(T_{\varsigma})
&<C_1 2^{\varsigma}\omega\le C_2\frac{r}{\log(r)}\log(r)=C_2 r;
\end{align*}
here we take into account the fact that all coefficients
$a(i)$, $b(i)$, $p(j)$, $q(j)$ are polynomials.

Let's estimate the time complexity of the calculation of $\sigma_t'$.
We must take into account the property of the complexity of integer
multiplication: $2M(2^{-1}m)\le M(m)$ (quasi-linear and polynomial functions satisfy
this property). Because at a tree node of recursive calls at level $i$
there are $C_3$ multiplications of $2^{\varsigma-i+1}$ numbers of length at most
$2^{i+1}\omega+2^{i}$, the following estimate of the number of operations
required to calculate $\sigma_t'$ holds:
\begin{align*}
Time(\sigma_t')
&\le C_3\sum_{i=1}^{\varsigma}{2^{\varsigma-i}M(2^{i+1}\omega+2^{i})}
< C_4\sum_{i=1}^{\varsigma}{2^{\varsigma-i}M(2^{i+2}\omega)}\\
&= C_4\sum_{i=1}^{\varsigma}{2^{\varsigma-i}M(2^{\varsigma-(\varsigma-(i+2))}\omega)}
\le C_5\sum_{i=1}^{\varsigma}{2^{\varsigma-i}2^{-\varsigma+(i+2)}M(2^{\varsigma}\omega)}\\
&\le C_6 \varsigma M(r)\le C_7\log(r)M(r)
\end{align*}
(here we use the inequality for $2^{\varsigma}\omega$ from the estimate of
$l(T_{\varsigma})$). Final division gives $O(M(r))$ operations.

Now we estimate the space complexity of the computation of $\sigma_t'$.
Because at a chain element at level $i$ the amount of
memory used for temporary variables is $C_8(2^{i+1}\omega+2^{i})$,
the amount of memory in all simultaneously existing recursive calls is
estimated as follows:
\begin{align*}
Space(\sigma_t')
\le\sum_{i=1}^{\varsigma}{C_8(2^{i+1}\omega+2^{i})}
\le C_{9}(2^{\varsigma}\omega+2^{\varsigma})
\le C_{10}r=O(r).
\end{align*}
\end{proof}
\end{lemma}
\begin{lemma}[2]
The time complexity of the binary splitting method for the calculation of
$\tau_t$ is bounded by $O(M(r)\log(r))$; its space complexity is bounded by $O(r)$.
\begin{proof}
The estimation of the computational complexity of $\tau_t$ is the same as
the estimation for $\sigma_t$ due to the fact that the inequalities in the proof of
lemma (1) are also suitable for $\tau_t$ (we can calculate the numerator and
denominator of $\sigma_t$ using the binary splitting method for products).
\end{proof}
\end{lemma}
We will calculate approximate values $P(r)^*$ with accuracy
$2^{-(n+1)}$ in accordance with Formula \eqref{eq:yah:PmukEquation}
using the following iterative process:
\begin{align}
&h_1(m)=\sigma_{k_1}^*,\notag\\
&\widehat h_i(m)=\sigma_{k_1-i+1}^*+\tau_{k_1-i+2}^* h_{i-1},
\quad i=1,\dots,k_1,
\label{eq:yah:hiScheme}
\\
&h_i(m)=\widehat h_i(m)+\varepsilon_i;\notag
\end{align}
for $i=k_1$ we suppose $P(r)^*=h_{k_1}(m)$. Here $m\ge r$ ($m$ will be
chosen later), $\sigma_i^*$ and $\tau_i^* $ are approximations of $\sigma_i$
and $\tau_i$ with accuracy $2 ^{-m}$.
The values $h_i(m)$ are obtained by discarding bits
$q_{m+1}q_{m+2}\dots q_{m+j}$ of numbers $\widehat h_i(m)$ after
the binary point starting with the $(m+1)$ bit:
\begin{align}
|\varepsilon_i|=|h_i(m)-\widehat h_i(m)|=
0.0\dots 0 q_{m+1}q_{m+2}\dots q_{m+j},
\label{eq:yah:varepsilonEquation}
\end{align}
and the sign of $\varepsilon_i$ is the same as the sign of $\widehat h_i(m)$
(it is clear that $|\varepsilon_i|<2^{-m}$).

Let's assume that the following conditions hold:
\begin{align}
b(i)\ge 2\thickspace\text{for all}\thickspace i,
\quad\frac{p(j)}{q(j)}\le 1\thickspace\text{for all}\thickspace j.
\label{eq:yah:ConditionsOnSeries}
\end{align}
We prove the following two lemmas.

\begin{lemma}[3]
For every $i\in 1..k_1$ 
\begin{align}
|h_i(m)|<(i+1)r_1 W.
\label{eq:yah:lemma1Inequality}
\end{align}
\begin{proof}
We apply induction on $j$ to $h_j(m)$. The induction base is $j=1$:
$|h_1(m)|\le\frac{1}{2}r_1 W+2^{-m}<2 r_1W$. The induction step is
$(j+1)\geq 2$:
\begin{align*}
|h_{j+1}(m)|
&=|\sigma_{k_1-(j+1)+1}^*+\tau_{k_1-(j+1)+2}^* h_{j}+\varepsilon_{j+1}|\\
&\le\frac{1}{2}r_1 W+(j+1)r_1 W+2^{-m}<
((j+1)+1)r_1 W.
\end{align*}
\end{proof}
\end{lemma}
\begin{lemma}[4]
The error of calculation of $h_{k_1}(m)$ according to scheme
\eqref{eq:yah:hiScheme} is estimated as
\begin{align*}
\Delta(k_1,m)<2^{-m}m k_1^{2}W.
\end{align*}
\begin{proof}
Let's denote
\begin{align*}
H_1=\sigma_{k_1},\quad
H_i=\sigma_{k_1-i+1} + \tau_{k_1-i+2}H_{i-1},\quad
\eta(i,m)=|h_i(m)-H_i|.
\end{align*}
We use the method of mathematical induction for $\eta(j,m)$
for $j$. The induction base is $j=1$:
\begin{align*}
\eta(1,m)=|h_1(m)-H_1|=|\sigma_{k_1}^*-\sigma_{k_1}|<2^{-m}.
\end{align*}
The induction step is $(j+1)\geq 2$:
\begin{align*}
\eta(j+1,m)
&=|\sigma_{k_1-(j+1)+1}^*+\tau_{k_1-(j+1)+2}^*h_{j}(m)+\varepsilon_{j+1}-\\
&\quad\sigma_{k_1-(j+1)+1}-\tau_{k_1-(j+1)+2}H_{j}|\\
&<|\tau_{\upsilon}^*h_{j}(m)-\tau_{\upsilon}h_{j}(m)+
\tau_{\upsilon}h_{j}(m)-\tau_{\upsilon}H_{j}|+2\cdot 2^{-m}\\
&\leq 2^{-m}h_{j}(m)+\tau_{\upsilon}\eta(j,m)+2\cdot 2^{-m}.
\end{align*}
Because of \eqref{eq:yah:lemma1Inequality} $|h_j(m)|<(j+1) r_1 W $ and,
by the induction hypothesis, $\eta(j, m)<2^{-m} mj^2 W $, we get
\begin{align*}
\eta(j+1,m)
&<2^{-m}(j+1)r_1 W+2^{-m} m j^2 W+2\cdot 2^{-m}\\
&<2^{-m}(j+1)m W+2^{-m} m (j+1)^2 W = 2^{-m}m(j+2)^2 W.
\end{align*}
From $\Delta(k_1,m)=\eta(k_1,m)$ we get the required inequality.
\end{proof}
\end{lemma}
Lemma 4 implies that to compute $P(r)^*$ with accuracy $2^{-(n+1)}$
it suffices to take $m$ such that the following holds
\begin{align}
m\ge(n+1)+\lceil \log(n+1)+2\log(r)+\log(W)\rceil,
\label{eq:yah:mRule}.
\end{align}
We denote the algorithm for the calculation of the hypergeometric series
which uses scheme \eqref{eq:yah:hiScheme} as {\itshape LinSpaceBinSplit}
(binary splitting method with linear space complexity).

\smallskip
{\bfseries
{\itshape Algorithm {\normalfont \itshape LinSpaceBinSplit}.}
The approximate value of the hypergeometric series.}\\
\emph{Input:} Record of the accuracy $2^{-n}$.\\
\emph{Output:} The approximate value of \eqref{eq:yah:SeriesDef}
with accuracy $2^{-n}$.\\
\emph{Description:}
\begin{enumerate}
\item[1)]{compute $r:=\mu(n+1)$; choose $k_1$ so that $2^{k_1}\ge r$;
compute $r_1:=\lceil\frac{r}{k_1}\rceil$; compute $W$;}
\item[2)]{find $m$ using Formula \eqref{eq:yah:mRule};}
\item[3)]{$h:=\sigma_{k_1}^*$ (using the usual binary splitting method
with accuracy $2^{-m}$);}
\item[4)]{make loops through $i$ from $2$ to $k_1$:
\begin{enumerate}
\item[a)]{calculate $v_1:=\sigma_{k_1-i+1}^*$ with accuracy $2^{-m}$
using the usual binary splitting method and $v_2:=\tau_{k_1-i+2}^*$ with
accuracy $2^{-m}$,}
\item[b)]{calculate the expression $\widehat h:=v_1+v_2 h$,}
\item[c)]{assign value $\widehat h$ rounded
in accordance with \eqref{eq:yah:varepsilonEquation} to $h$;}
\end{enumerate}
}
\item[5)]{write $h$ to output.}
\end{enumerate}
We estimate the time computational complexity of the algorithm
taking into account that $r$ and $m$ are linearly dependent on $n$:
\begin{itemize}
\item{$O(\log(n))$ computations of $\sigma_t$ gives
$O(M(n)\log(n)^2)$;}
\item{$O(\log(n))$ computations of $\tau_t$ gives
$O(M(n)\log(n)^2)$;}
\item{$O(\log(n))$ multiplications of numbers of the length $O(n)$ gives
$O(M(n)\log(n))$;}
\end{itemize}
in total we obtain $O(M(n)\log(n)^2)$ bit operations. The space
complexity of algorithm {\itshape LinSpaceBinSplit} is $O (n)$ because in
all calculations in this algorithm numbers of length $O(n)$
are used.

\begin{theorem}
The modified binary splitting algorithm for the calculation of 
the hypergeometric series, {\itshape LinSpaceBinSplit}, belongs to 
$Sch(FQLIN-TIME//LIN-SPACE)$.
\end {theorem}


\paragraph{Conclusion.}
In Tables \ref{yakhontov:tab1} and \ref{yakhontov:tab2} there are formulas
and series for the calculation of some frequently used constants
of mathematical analysis. These series converge linearly
(see \cite{yakhontov:bib:Haible, yakhontov:bib:Cheng}) and
they satisfy conditions \eqref{eq:yah:ConditionsOnSeries};
hence, to calculate them, we can use 
{\itshape LinSpaceBinSplit} and therefore these constants belong to
the set of constructive real numbers $Sch(FQLIN-TIME//LIN-SPACE)_{CF}$.
The algorithm {\itshape LinSpaceBinSplit} can also be used to calculate approximations
to many other constants and to the values of elementary functions
at rational points.
\begin{table}[t!]
\centering
\begin{minipage}[t]{128mm}\small\centering
\caption{\bf Formulas for evaluation of constants}\vspace{1mm}
\begin{tabular}{|l|l|l|}
\hline
{\bfseries Constant} & {\bfseries Formula} & {\bfseries Series}
\\
\hline
$e$
&
$e=\exp(1)$
&
$\exp(1)=2\sum_{i=0}^{\infty}{\frac{1}{2i!}}$
\\
\hline
$\pi$
&
$\pi=16\alpha-4\beta$,
&
$\arctg\left(\frac{1}{5}\right)=2\frac{1}{5}
\sum_{i=0}^{\infty}{(-1)^{i}\frac{1}{2(2i+1)5^{2i}}}$,
\\
&
$\alpha=\arctg{\frac{1}{5}}$,
&
$\arctg\left(\frac{1}{239}\right)=2\frac{1}{239}
\sum_{i=0}^{\infty}{(-1)^{i}\frac{1}{2(2i+1)239^{2i}}}$
\\
&
$\beta=\arctg{\frac{1}{239}}$
&
\\
\hline
$\zeta(3)$
&
&
$\sum_{i=0}^{\infty}{\frac{(-1)^i(205i^2+250i+77)((i+1)!)^5(i!)^5}
{2((2i+2)!)^5}}$\\
\hline
\end{tabular}
\label{yakhontov:tab1}
\end{minipage}
\end{table}
\begin{table}[t!]
\centering
\begin{minipage}[t]{128mm} \small \centering
\caption{\bf Series for calculations of constants}\vspace{1mm}
\begin{tabular}{|l|c|c|c|c|}
\hline
{\bfseries Constant} & $\mathbf{a(i)}$ & $\mathbf{b(i)}$ &
$\mathbf{p(j)}$ & $\mathbf{q(j)}$ \\
\hline
$e$ &
$1$ & $2$ & $1$ & $j$\\
\hline
$\pi$ & $1$
& $2(2i+1)$ & $-1$ & $5^2$ \\
& $1$
& $2(2i+1)$ & $-1$ & $239^2$ \\
\hline
$\zeta(3)$ &
$205i^2+250i+77$ & $2$ & $p(0)=1$, & $32(j+1)^5$\\
& & &
$p(j)=-j^5$ &\\
\hline
\end{tabular}
\label{yakhontov:tab2}
\end{minipage}
\end{table}

If we use the Schonhage--Strassen algorithm for integer multiplication with
a time complexity of $O(n\log(n)\log\log(n))$, then the time complexity of
{\itshape LinSpaceBinSplit} will be $O(n\log(n)^3\log\log(n))$;
when we use a simple recursive method for integer multiplication with a time
complexity of $O(n^{\log(3)})$, the time complexity of 
{\itshape LinSpaceBinSplit} will be $O(n^{\log(3)}\log(n)^2) $.

Let's note that the series of the constant $e$ converges with the rate
$2^{-O(n\log(n))}$, so the time complexity of the calculation of $e$
using {\itshape LinSpaceBinSplit} will be $O(M(n)\log(n))$ and
its space complexity will be $O\left(\frac{n}{\log(n)}\right)$.



\begin {thebibliography} {10}

\bibitem{yakhontov:bib:Schonhage}
   Schonhage~A., Grotefeld~A.\,F.\,W, Vetter E.
   \textit{Fast Algorithms. A Multitape Turing Machine Implementation.} //
   Germany: Brockhaus, 1994. 

\bibitem{yakhontov:bib:Haible}
   Haible B., Papanikolaou T.
   Fast multiple-precision evaluation of series of
   rational numbers // \textit{Proc. of the Third Intern.
   Symposium on Algorithmic Number Theory}. June 21--25, 1998.
   pp. 338--350.

\bibitem{yakhontov:bib:Cheng}
   Cheng H., Gergel B., Kim E., Zima E.
   Space-efficient evaluation of hypergeometric series //
   \textit{ACM SIGSAM Bull. Communications in Computer Algebra}. 2005.
   Vol. 39, No. 2. pp. 41--52.

\bibitem{yakhontov:bib:Ko}
   Ko K.
   \textit{Complexity Theory of Real Functions}. //
   Boston: Birkhauser, 1991.

\end{thebibliography}
\end{document}